# A high-mobility two-dimensional electron gas at the heteroepitaxial spinel/perovskite complex oxide interface of γ-Al$_2$O$_3$/SrTiO$_3$

Y. Z. Chen[1*], N. Bovet[2], F. Trier[1], D. V. Christensen[1], F. M. Qu[3], N. H. Andersen[4], T. Kasama[5], W. Zhang[1], R. Giraud[6,8], J. Dufouleur[6], T. S. Jespersen[7], J. R. Sun[3], A. Smith[1], J. Nygård[7], L. Lu[3], B. Büchner[6], B. G. Shen[3], S. Linderoth[1], and N. Pryds[1]

[1]*Department of Energy Conversion and Storage, Technical University of Denmark, Risø Campus, 4000 Roskilde, Denmark*

[2]*Nano-Science Center, Department of Chemistry, University of Copenhagen, 2100 Copenhagen, Denmark*

[3]*Institute of Physics, Chinese Academy of Sciences, 100190 Beijing, China*

[4]*Department of Physics, Technical University of Denmark, 2800 Lyngby, Denmark*

[5]*Center for Electron Nanoscopy, Technical University of Denmark, 2800 Lyngby, Denmark*

[6]*Leibniz Institute for Solid State and Materials Research, IFW Dresden, D-01171 Dresden, Germany*

[7]*Niels Bohr Institute, University of Copenhagen, 2100 Copenhagen, Denmark*

[8]*Laboratoire de Photonique et de Nanostructures-CNRS, Route de Nozay, 91460 Marcoussis, France*

**The discovery of two-dimensional electron gases (2DEGs) at the heterointerface between two insulating perovskite-type oxides, such as LaAlO$_3$ and SrTiO$_3$, provides opportunities for a new generation of all-oxide electronic and photonic**

---

[*] Correspondence and requests for materials should be addressed to Y.Z.C (Email: yunc@dtu.dk).

devices. However, significant improvement of the interfacial electron mobility beyond the current value of approximately 1,000 $cm^2V^{-1}s^{-1}$ (at low temperatures), remains a key challenge for fundamental as well as applied research of complex oxides. Here, we present a new type of 2DEG created at the heterointerface between $SrTiO_3$ and a spinel gamma-$Al_2O_3$ epitaxial film with excellent quality and compatible oxygen ions sublattices. This spinel/perovskite oxide heterointerface exhibits electron mobilities more than one order of magnitude higher than those of perovskite/perovskite oxide interfaces, and demonstrates unambiguous two-dimensional conduction character as revealed by the observation of quantum magnetoresistance oscillations. Furthermore, we find that the spinel/perovskite 2DEG results from interface-stabilized oxygen vacancies and is confined within a layer of 0.9 nm in proximity to the heterointerface. Our findings pave the way for studies of mesoscopic physics with complex oxides and design of high-mobility all-oxide electronic devices.

High-mobility 2DEGs confined in epitaxially grown semiconductor heterostructures form the basis of modern electronic and photonic devices, and have constituted the material basis for the development of quantum transport and mesoscopic physics, for example, the resultant discoveries of the integer and fractional quantum Hall effects[1,2]. Different from those in semiconductors, strongly correlated electrons in complex oxides with partially occupied *d*-orbitals give rise to a variety of extraordinary electronic properties, such as high-temperature superconductivity, colossal magnetoresistance, ferromagnetism, ferroelectricity, and multiferroicity. Therefore, the high-mobility 2DEGs at atomically engineered complex oxide interfaces not only show promise for multifunctional all-oxide devices with probably even richer behavior than that in bulk[3-10], but would also provide a wealth of opportunities to study mesoscopic physics with strongly correlated electrons confined in nanostructures. Nevertheless, this requires a large-enough electron mobility, so that the characteristic lengths of the system, such as the mean free path or the phase coherence length, become sizeable with respect to the typical dimension of quantum devices.

The enhancement of electron mobilities for complex oxide 2DEGs, however, meets formidable challenges. To date, these 2DEGs have been fabricated exclusively at oxide interfaces between perovskite bilayers[4], such as the (001)-oriented polar $LaAlO_3$ (LAO) films grown epitaxially on (001)-oriented nonpolar $SrTiO_3$ (STO) single crystals with a $TiO_2$-termination[3]. The two-dimensional electron mobility in these perovskite-type oxide interfaces is typically ~1000 $cm^2V^{-1}s^{-1}$ at 2 K[4,10], with a sheet carrier density, $n_s$, being $10^{13}$-$10^{14}$ $cm^{-2}$. This Hall mobility is still much lower than those for three-dimensional oxygen-deficient STO single crystals[11] and La-doped STO epitaxial films[12], amounting to $1.3 \times 10^4$ $cm^2V^{-1}s^{-1}$ and $3.2 \times 10^4$ $cm^2V^{-1}s^{-1}$, respectively. The 2DEGs at these perovskite-type oxide interfaces are suggested to result from electronic reconstructions due to a polar discontinuity at the interface[3]; however, mechanisms such as ion transfer across the interface and formation of defects have also been identified to play important roles on the transport properties[13,14]. Harnessing the impurities and defects at these polar complex oxide interfaces remains elusive[15]. Despite deliberate efforts, the highest electron Hall mobility observed in the LAO/STO-based oxide interfaces is limited to the order of 5000 $cm^2V^{-1}s^{-1}$ at 2 K[16,17]. Besides interface polarity, we have recently found that chemical redox reactions at the oxide interface between STO single crystals and other complex oxides containing Al, Ti, Zr, and Hf elements can provide an alternative approach to creating 2DEGs in complex oxide heterostructures[18]. Nevertheless, establishing electron confinement with increased carrier mobilities in STO-based heterointerfaces remains a challenge[18].

Here we present a novel 2DEG with electron Hall mobilities as large as $1.4 \times 10^5$ $cm^2V^{-1}s^{-1}$ and $n_s$ as high as $3.7 \times 10^{14}$ $cm^{-2}$ at 2 K by creating a spinel/perovskite complex oxide interface between epitaxial alumina ($Al_2O_3$) films and STO single crystals (Fig. 1a). To our knowledge, it is the first time that complex oxide interfaces based on STO are found to exhibit carrier mobilities larger than any yet reported for either electron-doped STO single crystals[11] or optimized epitaxial doped-STO films[12]. Moreover, such a high mobility opens the door to the design of mesoscopic quantum devices based on complex oxides.

## Results

**Sub-unit-cell layer-by-layer two-dimensional growth of γ-$Al_2O_3$ films on $TiO_2$-terminated STO.** $Al_2O_3$ is a widely used oxide and is also one of the best insulating materials in nature with a band gap normally above 8.0 eV. The synthesis of nanoscale $Al_2O_3$ usually results in $\gamma$-$Al_2O_3$ with a spinel-type structure, rather than the common $\alpha$-$Al_2O_3$ with a corundum structure because the $\gamma$-$Al_2O_3$ has a lower surface energy than $\alpha$-$Al_2O_3$[19]. Remarkably, as illustrated in Fig. 1b-d, despite differences in cation sublattices, the oxygen sublattice of the spinel $\gamma$-$Al_2O_3$ matches closely with that of the perovskite STO, since the lattice parameter of $\gamma$-$Al_2O_3$ is twice that of STO ($a_{STO}$ = 3.905 Å, $a_{\gamma-Al2O3}$ = 7.911 Å[20], lattice mismatch of 1.2%). Such an excellent lattice match between oxygen sublattices, together with the low surface energy of $\gamma$-$Al_2O_3$, makes it compatible to grow epitaxially GAO/STO (GAO/STO) spinel/perovskite heterostructures in a persistent two-dimensional layer-by-layer growth mode (Supplementary Information, Fig. S1). Figure 1e shows typical intensity oscillations of the reflection high-energy electron diffraction (RHEED) pattern during the growth of a 3 unit cells (uc) $\gamma$-$Al_2O_3$ film at a growth temperature of 600 ºC. For the epitaxial growth of ionic oxides, when all film components are supplied simultaneously, the oscillation period corresponds to the minimum unit of the chemical composition needed to ensure charge neutrality[21-23]. For $\gamma$-$Al_2O_3$ grown along the (001) direction, one intensity oscillation corresponds to the growth of one quarter unit cell film (Fig. 1e), since the $\gamma$-$Al_2O_3$ unit cell consists of four neutral "$AlO_x$" sub-unit cells with an interlayer distance of about 0.2 nm. Similar sub-unit-cell layer-by-layer film growth has been observed in the epitaxial growth of spinel magnetite ($Fe_3O_4$)[24]. The persistent layer-by-layer two-dimensional film growth results in a high quality cubic-on-cubic GAO/STO epitaxial heterointerface with no obvious dislocations as confirmed by scanning transmission electron microscopy (STEM) (Fig.1f and g).

**Electrical transport properties of GAO/STO heterointerfaces.** The investigation of conductivity in our GAO/STO heterostructures shows that the interface between the two insulators can become metallic with electrons as the dominant charge carriers (Supplementary Information, Fig. S2). Note that, under the condition of our film growth, the bare STO substrate remains highly insulating without film deposition. More strikingly, 2DEGs with extremely high Hall electron mobilities are obtained when the γ-$Al_2O_3$ film is grown at an oxygen background pressure of $10^{-4}$ mbar and a growth temperature of 600 ºC (Fig. 2a-c). As shown in Fig. 2, the interfacial conduction depends critically on the thickness, $d$, of the γ-$Al_2O_3$ film. The heterointerface changes from highly insulating to metallic when $d$ is above a threshold thickness of approximately 2 uc (Fig. 2d and e). At $d$=2 uc, the interface shows a sheet resistance, $R_s$, and a carrier density, $n_s$, in the order of 10 kΩ/□ and $2.3×10^{13}$ cm$^{-2}$ at $T$=300 K, respectively, similar to the perovskite-type LAO/STO interface[4,6,7,9,14,17]. Remarkably, we find a striking $R_s$ decrease of about three orders in magnitude and a Hall mobility as high as $\mu_{Hall} \approx 1.1×10^4$ cm$^2$V$^{-1}$s$^{-1}$ at $T$=2 K in the spinel/perovskite GAO/STO interface. By carefully controlling the film growth down to a sub-unit-cell level, a great $R_s$ decrease of approximately four orders in magnitude is observed at $d$=2.5 uc, which is accompanied by the presence of nonlinear Hall resistance with respect to magnetic fields at temperatures below 100 K (Supplementary Information, Fig. S3). A linear fitting to the low-field Hall resistance gives rise to an impressive $\mu_{Hall}$ of approximately $1.4×10^5$ cm$^2$V$^{-1}$s$^{-1}$ with a $n_s$ of $3.7×10^{14}$ cm$^{-2}$ at 2 K, which are consistent with those obtained by fitting the entire nonlinear Hall effect within a two-band model (Supplementary Information, Fig. S3). Note that the high mobility 2DEGs with $\mu_{Hall} \geq 10^4$ cm$^2$V$^{-1}$s$^{-1}$ at $T$=2 K are only detected in the thickness range of 2 uc $\leq d <$ 3 uc.

Further increasing $d$ deteriorates the electron mobility to less than 1000 cm$^2$V$^{-1}$s$^{-1}$, probably due to the significant outward diffusion of the Ti-cations across the interface as observed by electron energy loss spectroscopy (EELS) (Supplementary Information, Fig. S4).

**Two-dimensional quantum oscillations of the high-mobility conduction at GAO/STO heterointerfaces.** The two-dimensional nature of the conduction in our spinel/perovskite heterostructures is confirmed by angle-dependent Shubnikov-de Haas (SdH) quantum oscillations, which are superimposed on a huge background of positive magnetoresistance (Fig. 3a). After subtracting the magnetoresistance background, the SdH oscillations become apparent (Fig. 3b) and the extrema positions show a cosine dependence with the angle $\theta$ between the magnetic field and the surface normal (Fig. 3c). This reveals the two-dimensional nature of the electron gas formed at our GAO/STO interfaces. Besides, the absence of oscillations at $\theta = 90°$ further confirms that the spatial width of the 2DEG is smaller than at least the cyclotron radius at 15 T, the typical value of which is below 10 nm for our heterostructures. Moreover, the angular dependence of the SdH oscillations measured at high magnetic fields suggests a multiple-subband contribution to charge transport. For instance, an extra feature is observed at $\theta=50º$ with $B\cos\theta = 7.2$ T, which may result from a π shift of the oscillations due to a spin-split band. Such a phase shift has been observed in the high-mobility 2DEG of GaN/AlGaN interfaces when the Zeeman energy (depending on the total $B$) and the cyclotron energy (depending on the perpendicular component of $B$) are equal[25]. To confirm the high mobility achieved in our GAO/STO 2DEGs, we increased the visibility of the SdH oscillations by cooling one sample ($d$=2.25 uc) down to 22 mK in a dilution refrigerator. Ultra-low noise measurements allow us to evidence the oscillations down to about 1 T (Fig. 4a), which directly shows that the quantum mobility extracted from the SdH oscillations, $\mu_{SdH}$, is in the range of $10^4$ cm$^2$V$^{-1}$s$^{-1}$, as inferred from the onset of oscillations. Importantly, the low-field dependence of the SdH oscillations reveals the typical behavior due to a single band. According to theory[26], the oscillations amplitude $\Delta R_{xx}$ can be described as:

$$\Delta R_{xx} = 4R_0 e^{-\alpha T_D} \alpha T / \sinh(\alpha T)$$

Where, $\alpha = 2\pi^2 k_B / \hbar \omega_c$, $\omega_c = eB/m^*$ is the cyclotron frequency, $m^*$ is the carrier effective mass, $k_B$ is Boltzmann's constant and $\hbar$ is Planck's constant divided by $2\pi$. $R_0$ is the classical resistance in zero field. $T_D = \hbar / 2\pi k_B \tau$ is the Dingle temperature, $\tau$ is the total scattering time. At a fixed magnetic field, $m^*$ can be deduced by fitting the temperature-dependent oscillation amplitude with

$\Delta R_{xx}(T)/\Delta R_{xx}(T_0) = T \sinh(\alpha T_0)/T_0 \sinh(\alpha T)$ ($T_0$=22 mK). As shown in Fig. 4b for $B$=2.04 T, the fit leads to an effective mass of $m^*$=(1.22±0.03) $m_e$ ($m_e$ is the bare electron mass), consistent with those reported for other STO-based heterostructures[16, 27-30]. At a fixed temperature, $T_D$ or $\tau$ can be deduced from the slope of the Dingle plot, i.e., $\ln[\Delta R_{xx} \sinh(\alpha T)/4R_0 \alpha T]$ versus $1/B$ (Fig. 4c for $T$=200 mK), which gives a $\tau$ =4.96×10$^{-12}$ s or $T_D$= 0.24 K, corresponding to a quantum mobility $\mu_{SdH}= e\tau/m^*$ of 7.2 ×10$^3$ cm$^2$V$^{-1}$s$^{-1}$. Such an unprecedented high $\mu_{SdH}$ in our GAO/STO 2DEGs is more than 1 order of magnitude higher than those observed in the perovskite/perovskite LAO/STO[16,17,27] and GaTiO$_3$/STO[30] heterostructures, which are typically below 300 cm$^2$V$^{-1}$s$^{-1}$. Note that the difference between $\mu_{Hall}$ and $\mu_{SdH}$ in our GAO/STO heterostructures could come from a different scattering time (*i.e.* the transport scattering time and the total scattering time, respectively), which has also been reported in the LAO/STO[16, 17, 27] and $\delta$-doped STO heterostructures[28,29], as well as the GaAs/AlGaAs heterostructures[31]. In short, the SdH measurements support the formation of high-mobility 2DEGs at our spinel/perovskite heterointerfaces (see also Supplementary Information, Fig. S5).

**Spatial confinement of the 2DEG at the GAO/STO heterointerface.** To determine the origin and depth-profile for the conduction in the GAO/STO heterostructures, angle resolved X-ray photoelectron spectroscopy (XPS) measurements are performed. We find that the electrons are exclusively accumulated on the otherwise empty 3$d$ shell of $Ti^{4+}$ on the STO side. The most remarkable XPS result is that the $Ti^{3+}$ signal in GAO/STO heterointerfaces shows strong dependence on the photoelectrons detection angle, $\varphi$, with respect to the surface normal. An increase of the $Ti^{3+}$ signal with increasing $\varphi$, as shown in Fig. 5a, is clearly detected for $d$=2.5 uc with the highest Hall mobility. This further confirms that the conduction in our GAO/STO heterointerface is highly confined at the interface region. To make more quantitative analyses, we assume a simple case that the 2DEG extends from the interface to a depth, $t$, into the STO substrate[32]. The interface region is further assumed to be stoichiometric, sharp and characterized by a constant fraction, $p$, of $Ti^{3+}$ per STO unit cell. Taking into account the attenuation length of photoelectrons, the ratio of $Ti^{3+}$ to $Ti^{4+}$ signal, $I(Ti^{3+})/I(Ti^{4+})$, as a function of $\varphi$ can be calculated as[32]:

$$\frac{I(Ti^{3+})}{I(Ti^{4+})} = \frac{p[1-\exp(-t/\lambda\cos\varphi)]}{1-p[1-\exp(-t/\lambda\cos\varphi)]}$$

Where, $\lambda$ is the electron escape depth in STO. According to the NIST database[33], $\lambda$ is approximately 2.2 nm for our setup. As shown in Fig. 5b, the best fitting of the experimental $I(Ti^{3+})/I(Ti^{4+})$ ratios gives a $p \approx 0.31$, which equals to a $n_s \approx 2.1 \times 10^{14}$ cm$^{-2}$, and a $t$ of 0.9 nm. Therefore, the electrons at our GAO/STO heterointerface are strongly confined within approximately the first 2 uc of STO surface in proximity to the interface. Note that the $n_s$ deduced here is slightly lower than that obtained from Hall data (Fig. 2c). This could be due to the presence of outward diffusion of the Ti-cations into alumina films, where $Ti^{4+}$ is the dominant component (Supplementary Information, Fig. S4). Such concern is also consistent with the fact that the out-diffused Ti is found to have a negligible contribution to the measured interface conduction. For example, the interface conduction remains unaffected when the capping alumina film is etched away by a 4M aqueous NaOH solution. This strongly suggests that the effective charge carriers are mainly located on the STO side.

## Discussion

Since each layer of the (001)-oriented GAO/STO heterointerface is nominally charge neutral, the polar discontinuity induced electronic reconstruction as expected in the LAO/STO interface[3] is ruled out here. The presence of $Ti^{3+}$ is probably a signature of the formation of oxygen vacancies on the STO side. This scenario is consistent with the fact that the interfacial conductivity can be completely removed when the $Ti^{3+}$ content is significantly suppressed by suitable annealing in 1 bar pure $O_2$ at a temperature higher than 200 ºC (Supplementary Information, Fig. S6). Such an oxygen-vacancy-dominated 2DEG is expected to be formed as a consequence of chemical redox reactions occurring on the STO surface during the film growth of γ-$Al_2O_3$, analogous to what has been observed in metallic amorphous STO-based heterostructures grown at room temperature[18]. Note that the 2DEG at the crystalline GAO/STO heterointerface is formed at a high temperature of 600 ºC, where the oxygen ions in STO are already highly mobile. This is normally expected to level out any difference in the depth profile of oxygen distribution in STO[18, 34]. However, this is not the case in the crystalline GAO/STO heterostructures as inferred from both Fig. 3 and Fig. 5. Moreover, the conduction at the interface of thick films, for example at $d$=8 uc, can survive the annealing at 300 ºC for 24 hours in 1 bar pure $O_2$ with only negligible changes in the conductivity (Supplementary Information, Fig. S6). These features strongly suggest that the oxygen vacancies and the 2DEGs are stabilized by an interface effect, such as by the formation of a space charge region near the heterointerface. It is worth noting that an inherent oxygen ion deficiency has been observed at the grain boundary of STO bicrystals[35], where a considerable electron accumulation has also been predicted if the barrier height of the grain boundary is deliberately controlled[36]. The high electron mobility of STO-based oxide materials at low temperatures is generally related to the polarization shielding of the ionized defect scattering centers driven by the large dielectric constant of STO[37]. The higher mobility of our spinel/perovskite oxide interface compared to the perovskite-type oxide heterointerface may be due to the better lattice match and thereby a more perfect structure and well-defined interface. Though further investigations are needed to reveal how the interface properties increase the mobility and the associated strong suppression of the defect and impurity scattering, our results strongly suggest that defect engineering of oxygen vacancies is crucial for the high mobility of 2DEGs confined at the interface between complex oxides.

In conclusion, we have demonstrated that high-mobility 2DEGs with clear quantum magnetoresistance oscillations and strong spatial confinement can be created at well-defined spinel/perovskite GAO/STO oxide interfaces. The strongly spatial confinement of charge carriers achieved directly in the as-deposited spinel/perovskite oxide heterostructures without any post annealing provides the possibility to fabricate multilayers of complex oxides with several 2DEGs. Furthermore, by combining two of the largest groups of oxides, plenty of new physical properties, for instance, interfacial magnetism[6] and superconductivity[7] as observed in the perovskite-type LAO/STO interface, may be found at the GAO/STO heterointerface. Finally, with a large enhancement of the electron mobility, the GAO/STO heterointerface probably enables the design of mesoscopic quantum devices based on complex oxide 2DEGs and opens new avenues for oxide nanoelectronics and mesoscopic physics.

## Methods

**Sample growth.** The γ-Al$_2$O$_3$ thin films were grown by pulsed laser deposition (PLD)[38] using a KrF laser (λ = 248 nm) with a repetition rate of 1 Hz and laser fluence of 1.5 J cm$^{-2}$. The target-substrate distance was fixed at 5.6 cm. Commercial α-Al$_2$O$_3$ single crystals were used as targets. Singly TiO$_2$-terminated (001) STO crystals with a size of 5×5×0.5 mm$^3$ were used as substrates. Note that the TiO$_2$-termination of our substrates is obtained by chemical etching using HCL-HNO$_3$ as acidic solution[38], which is found to produce less defects on the STO surface compared to the conventional buffered HF etch method[39,40]. The film growth process was monitored by *in-situ* high pressure RHEED. During deposition, the oxygen pressure was fixed at 10$^{-4}$ mbar with the deposition temperature changing from room temperature (20 °C) to 700 °C. After film deposition, the samples were cooled down to room temperature at the deposition pressure. The film thickness was determined by both RHEED oscillations and X-ray reflectivity measurements.

**Electrical transport measurement.** The transport properties of the buried interface were measured using a four-probe Van der Pauw method with ultrasonically wire-bonded aluminum wires as electrodes, placed at the corners of the square sample. The temperature dependent electrical transport and Hall-effect measurements were performed in a CRYOGENIC cryogen free measurement system with the temperature ranging from 300 K down to 2 K and magnetic fields up to 16 T. To confirm the carrier density and mobility, some Hall-bar patterned samples were also measured, which were prepared directly through a mechanical mask[18]. Note that the use of a mechanical mask at deposition temperatures higher than 500 °C may have a deleterious effect on the carrier mobility, since the high oxygen ion diffusion can unintentionally disturb the oxygen equilibrium for realizing high mobility. The angle-dependent SdH measurements were performed in a sorption-pumped $^3$He cryostat with standard lock-in technique at 0.3 K with magnetic fields up to 15 T by changing the angles manually. The temperature-dependent SdH measurements were performed in a dilution refrigerator with a base temperature of 22 mK and an improved temperature stability, using ultra-low noise electronics. During all the transport measurements, the applied currents were within 1-10 μA (for AC current, the frequency was 327Hz). Special care was taken to avoid heating effect.

**X-ray photoelectron spectroscopy (XPS) measurement.** The XPS measurements were performed in a Kratos Axis Ultra$^{DLD}$ instrument, using a monochromatic Al Kα X-ray source with photon energy of 1486.6 eV. This leads to a kinetic energy of Ti 2*p* electrons of roughly 1025 eV. According to the NIST database[33], the electron escape depth is approximately 22 Å in STO at this kinetic energy. The pass energy used for the high resolution scan was 20 eV. The detection angle of the electrons was varied between 0° and 60° with respect to the sample normal. For analyzing the Ti 2$p_{3/2}$ peaks (Ti$^{4+}$ is at a binding energy of 459.5 eV, whereas the Ti$^{3+}$ is 1.6 eV ± 0.1 eV lower), a Shirley background was subtracted and the spectra were normalized to the total area below the Ti peaks ([Ti] = [Ti$^{4+}$] + [Ti$^{3+}$] = 100%).

**Scanning transmission electron microscopy (STEM) and electron energy-loss spectroscopy (EELS) measurements.** Aberration-corrected STEM measurements were performed by an FEI Titan 80-300ST

TEM equipped with a high brightness Shottky emitter (XFEG) and a Gatan Image Filter (Tridiem). High angle annular dark field (HAADF) images were acquired at 300 kV, where the probe size, convergence angle and HAADF collection angle were 0.8–1 Å, 20 mrad and 46-291 mrad, respectively. For EELS in the STEM, an accelerating voltage of 120 kV (probe size of 1.5-2.0 Å) was used to reduce knock-on damage to the specimen. The energy resolution of EELS was ~0.9 eV. Spectrum imaging was used to collect spectra across the interface. We typically recorded the spectrum images consisting of 40 ten-analysis-points lines (i.e. $10 \times 40$ pixel) parallel to the interface and acquired each line by an increment of 0.28 nm. Each spectrum was obtained at a dispersion of 0.1 eV for 0.2-0.4 sec. Then the spectra along the lines were summed after removing the spectra from beam damaged regions according to the HAADF contrast to increase signal/background ratio.

## Acknowledgements
We thank J. Fleig, F. W. Poulsen, N. Bonanos, S. Stemmer, and Y. Q. Li for helpful discussions. We also thank K. Thydén, Z. I. Balogh, J. W. Andreasen, E. Johnson, Y. Zhao, X. Tang, W.W. Gao, N. Y. Wu, J. Geyti, K. V. Hansen, K. Engelbrecht, and L. Theil Kuhn for their help.

## Author Contributions
Y.Z.C. concept design, film growth, transport measurements, data analysis, interpretation and writing of the manuscript. The contributions of other authors are as follows. Concept design: N.P., S.L.; XPS measurements and analysis: N.B.; Transport measurements and analysis: F.T., D.V.C., N.H.A., T.S.J.; SdH measurements and analysis: F.M.Q, R.G., J.D.; STEM and EELS measurements and analysis: T.K.; HRTEM measurements and analysis: W.Z.; Data discussion: J.R.S., A.S., J.N., L.L., B. B., B.G.S. All authors extensively discussed the results and the manuscript.

## Additional information
**Supplementary information** accompanies this paper on www.nature.com/naturecommunications

**Competing financial interests:** The authors declare no competing financial interests.

**Reprints and permissions** information is available online at http://npg.nature.com/reprintsandpermissions.

**Figure legends:**

**Figure 1 High-mobility 2DEGs at epitaxial spinel/perovskite GAO/STO interfaces. a**, A sketch of the heterostructure. **b**, Oxygen sublattices as the backbone to build the spinel/perovskite heterostructure. The compatibility in oxygen sublattices of a γ-$Al_2O_3$ surface and the $TiO_2$-terminated STO surface is shown in **c** and **d**, respectively. Note that the tetrahedral cation sites in γ-$Al_2O_3$ are not shown. **e**, Typical RHEED intensity oscillations for the growth of a 3-uc γ-$Al_2O_3$ on STO in a sub-unit-cell layer-by-layer mode. **f**, High-angle annular dark field (HAADF) STEM image of the epitaxial GAO/STO interface. Scale bar, 1 nm. Sr ions are brightest, followed by Ti. The faintly visible Al elements can be determined by the averaged line profiles across the interface shown in **g**. A well developed $TiO_2$-$AlO_x$ heterointerface is defined.

**Figure 2 Thickness-dependent electronic properties of the GAO/STO interface. a-c**, Temperature dependence of sheet resistance, $R_s$, carrier density, $n_s$, and low-field electron Hall mobility, $\mu_{Hall}$, for the interface conduction at different film thicknesses. **d** and **e**, Thickness dependence of the sheet conductance, $\sigma_s$, and $n_s$ measured at 300 K. High-mobility 2DEGs are obtained at a thickness range of 2 uc ≤ $d$ < 3 uc. The lines are guides to the eye.

**Figure 3 Two-dimensional quantum oscillations of the conduction at GAO/STO interfaces. a**, Longitudinal resistance, $R_{xx}$, as a function of magnetic field with visible SdH oscillations (arrowheads) under different tilt angle, $\theta$, at 0.3 K for the $d$=2.5 uc sample. **b** and **c**, Amplitude of the SdH oscillations, $\Delta R_{xx}$, under different $\theta$ versus the reciprocal total magnetic field and the reciprocal perpendicular magnetic field component, respectively. The SdH oscillations depend mainly on the reciprocal perpendicular magnetic field component, particularly in the $\theta$ angle of 0º-33º, which suggests a two-dimensional conduction nature of the GAO/STO interface.

**Figure 4 Small-field and low-temperature behaviour of the SdH oscillations. a**, Temperature dependence of the SdH oscillations at $\theta=0°$ for the $d$=2.25 uc sample. **b**, Temperature dependence of the scaled oscillation amplitude at $B$=2.04 T, giving a carrier effective mass of 1.22 $m_e$. **c**, Dingle plot of the SdH oscillations at 200 mK, giving a total scattering time $\tau$=4.96 ×$10^{-12}$ s, a related Dingle temperature $T_D$= 0.24 K, and a consequent quantum mobility $\mu_{SdH}$ = 7.2 ×$10^3$ $cm^2V^{-1}s^{-1}$.

**Figure 5 Spatial confinement of the 2DEG at the GAO/STO heterointerface determined by angle-resolved XPS. a**, The Ti $2p_{3/2}$ XPS spectra at various emission angles $\varphi$ for the $d$=2.5 uc sample. **b**, The angle dependence of the ratio of $Ti^{3+}$ to $Ti^{4+}$ signal, $I(Ti^{3+})/I(Ti^{4+})$, indicates a strong confinement of the conduction layer within 0.9 nm. Error bars indicate standard deviations, ±20%, for experimental values.

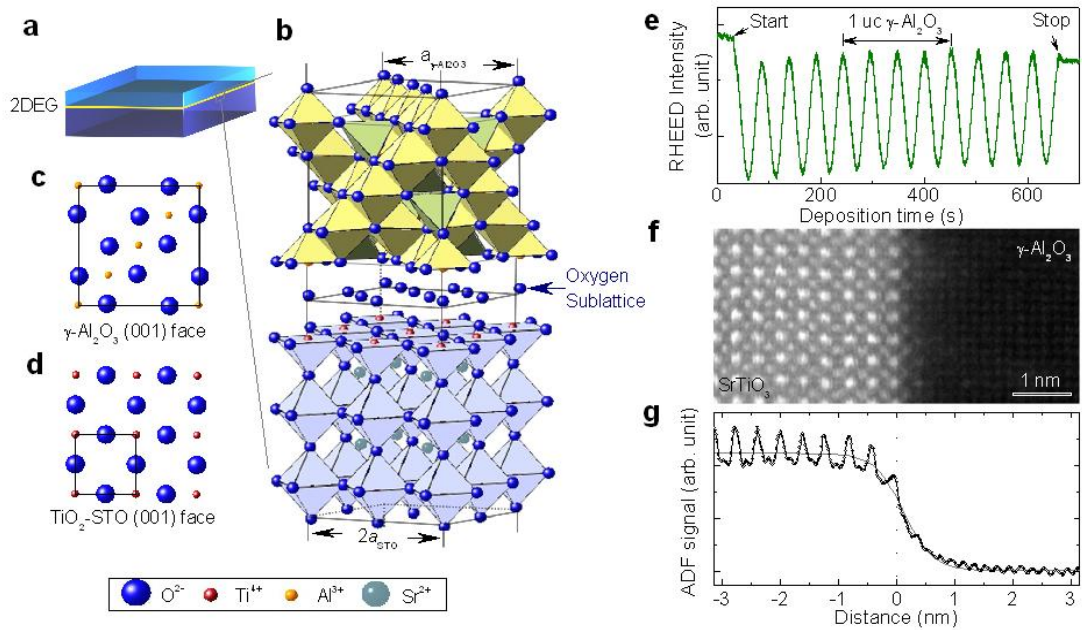

Fig.1 Y. Z. Chen *et al.*

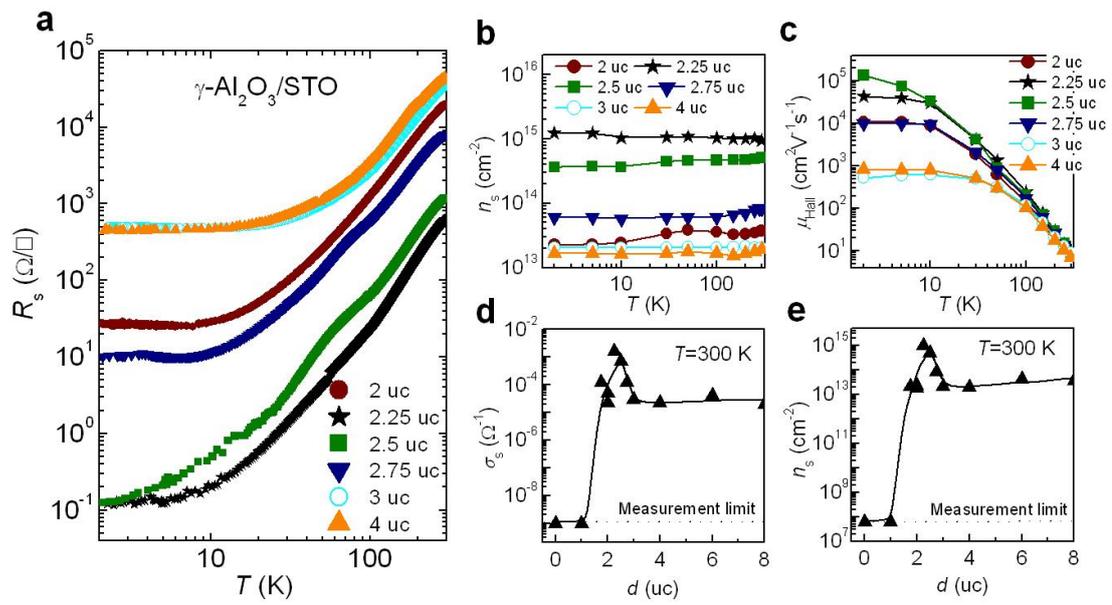

Fig. 2 Y.Z. Chen *et al.*

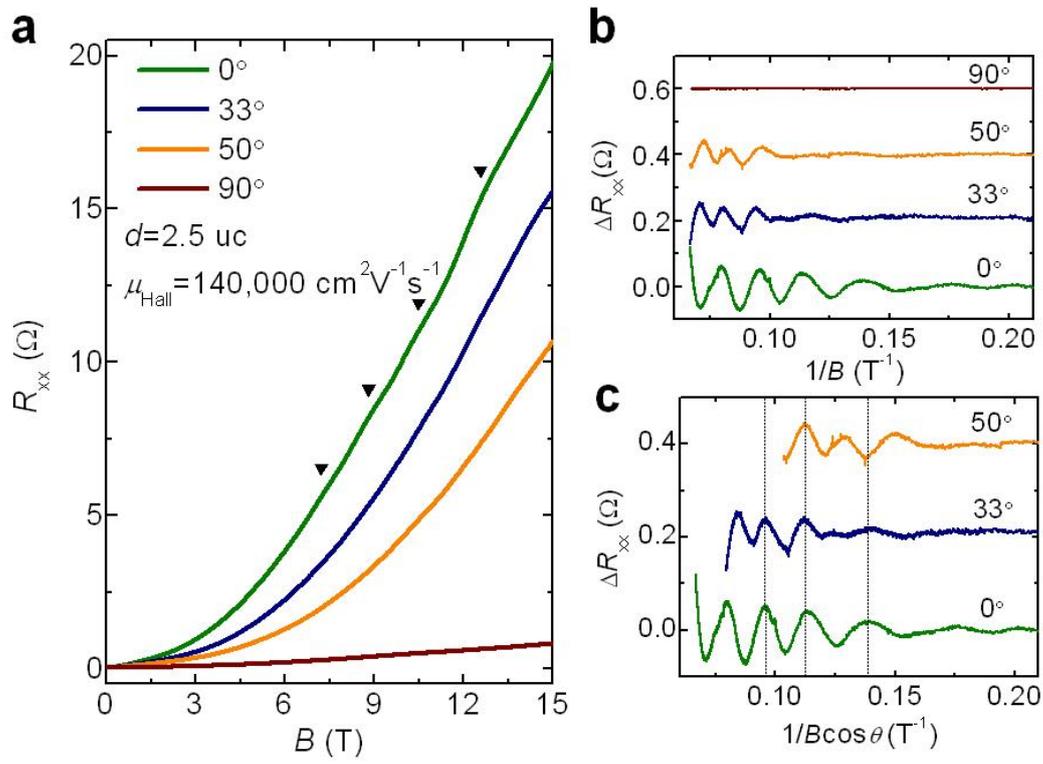

Fig. 3 Y. Z. Chen *et al.*

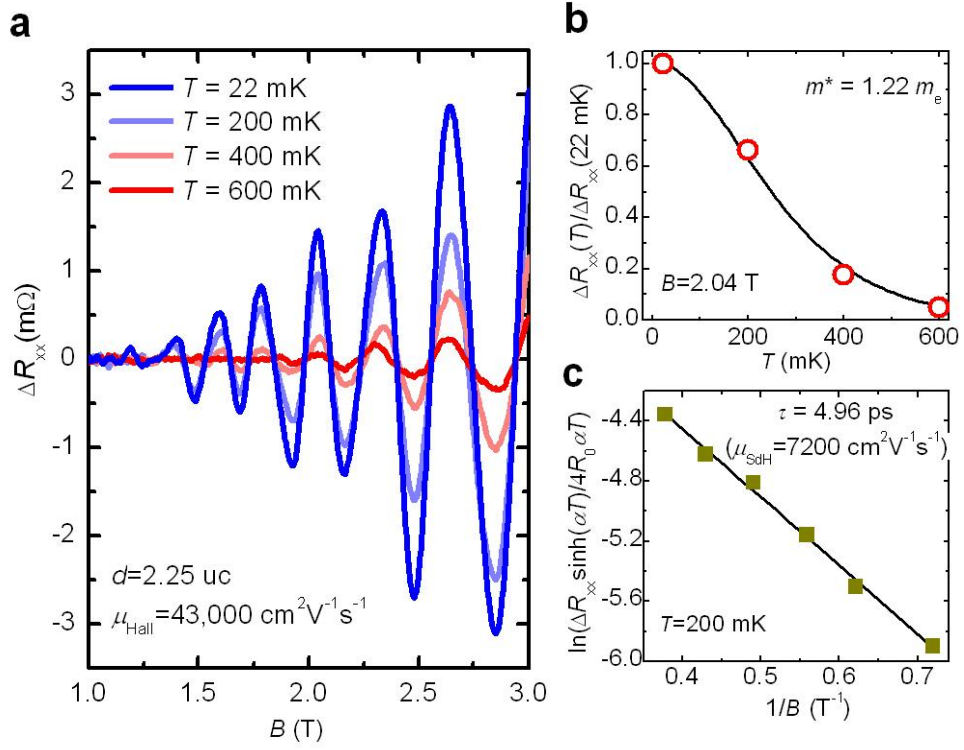

Fig. 4 Y. Z. Chen *et al.*

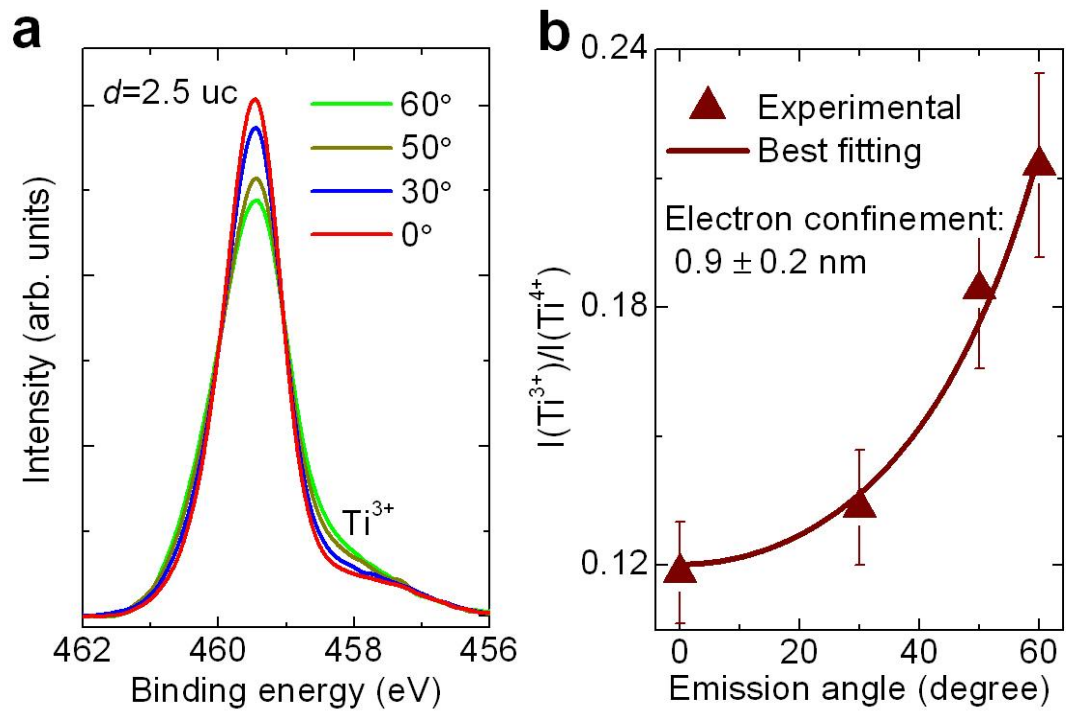

Fig. 5 Y. Z. Chen *et al.*